\documentclass
[superscriptaddress,secnumarabic,amssymb,amsmath,nobibnotes,aps,prd,showkeys,showpacs,nofootinbib,onecolumn,notitlepage]{revtex4}%
\usepackage{graphicx}
\usepackage{epsf}
\usepackage{bm}
\usepackage{amsmath}
\usepackage{amsfonts}
\usepackage{amssymb}
\usepackage{epstopdf}%
\providecommand{\U}[1]{\protect\rule{.1in}{.1in}}
\newcommand{\be}{\begin{equation}}
\newcommand{\ee}{\end{equation}}

\newcommand{\mincir}{\raise
-3.truept\hbox{\rlap{\hbox{$\sim$}}\raise4.truept\hbox{$<$}\ }}
\newcommand{\magcir}{\raise
-3.truept\hbox{\rlap{\hbox{$\sim$}}\raise4.truept\hbox{$>$}\ }}

\ifx\pdfoutput\relax\let\pdfoutput=\undefined\fi
\newcount\msipdfoutput
\ifx\pdfoutput\undefined\else
\ifcase\pdfoutput\else
\ifx\paperwidth\undefined\else
\ifdim\paperheight=0pt\relax\else\pdfpageheight\paperheight\fi
\ifdim\paperwidth=0pt\relax\else\pdfpagewidth\paperwidth\fi
\fi\fi\fi
\begin{document}
\title{(Compactified) black branes in four dimensional $f(R)$-gravity}
\author{N. Dimakis}
\email{nsdimakis@gmail.com}
\affiliation{Instituto de Ciencias F\'{\i}sicas y Matem\'{a}ticas, Universidad Austral de
Chile, Valdivia, Chile}
\author{Alex Giacomini}
\email{alexgiacomini@uach.cl}
\affiliation{Instituto de Ciencias F\'{\i}sicas y Matem\'{a}ticas, Universidad Austral de
Chile, Valdivia, Chile}
\author{Andronikos Paliathanasis}
\email{anpaliat@phys.uoa.gr}
\affiliation{Instituto de Ciencias F\'{\i}sicas y Matem\'{a}ticas, Universidad Austral de
Chile, Valdivia, Chile}
\affiliation{Institute of Systems Science, Durban University of Technology, PO Box 1334,
Durban 4000, Republic of South Africa}

\begin{abstract}
A new family of analytical solutions in a four dimensional static spacetime is
presented for $f\left(  R\right)  $-gravity. In contrast to General
Relativity, we find that a non trivial black brane/string solution is supported
in vacuum power law $f\left(  R\right)  $-gravity for appropriate values of
the parameters characterizing the model and when axisymmetry is introduced in
the line element. For the aforementioned solution, we perform a brief
investigation over its basic thermodynamic quantities.

\end{abstract}
\keywords{Exact solution; Black branes; Integrability; f(R)-gravity;}
\pacs{98.80.-k, 95.35.+d, 95.36.+x}
\maketitle
\date{\today}

\section{Introduction}

Modified theories of gravity have drawn the attention of the scientific
community because of the geometric mechanics that they provide to describe
various phenomena in nature. In this concept, new geometrodynamical degrees of
freedom are introduced in the gravitational field equations in such a way so
as to modify Einstein's General Relativity (GR). These additional terms can
have either theoretical or phenomenological origin \cite{clifton}. Among the
various\ proposed modified theories of gravity, $f\left(  R\right)  $-gravity
\cite{Buda} has been the main subject of study in various works over different
areas of gravitational physics.

$f\left(  R\right)  $-gravity is a fourth-order theory, where the the action
integral involves a function of the spacetime scalar curvature $R$. General
Relativity, with or without cosmological constant, is the special limit of
$f\left(  R\right)  $-gravity when the theory becomes of second-order; that
is, when $f$ is a linear function of the Ricci scalar. It belongs to a more
general family of theories that take into account curvature terms in order to
modify the Einstein-Hiblert action \cite{cur1,cur4,cur2,cur3}. The
gravitational field equations of $f\left(  R\right)  $-gravity are dynamically
equivalent to that of O'Hanlon theory \cite{Hanlon}, where a Lagrange
multiplier is introduced so as to reduce the order of the theory by increasing
the number of degrees of freedom through the introduction of a scalar field
\cite{Sotiriou,defelice}. This scalar field is nonminimally coupled to gravity
and recovers Brans-Dicke theory \cite{Brans} with a zero Brans-Dicke
parameter.\ Hence, $f\left(  R\right)  $-gravity is also related to families
of Horndeski theories \cite{hor}, which means that it is free of
Ostrogradsky's instabilities \cite{wood1,wood2}. For a recent discussion on
the correspondence among $f\left(  R\right)  $-gravity and other theories
through the various frames, together with relevant implications on
conservation laws, see \cite{san1, san2}.

As we mentioned before, the applications of $f\left(  R\right)  $-gravity in
gravitational physics cover various subjects. As far as cosmology is
concerned, the theory is used both to model the inflationary phase of the
universe \cite{star,inf1,inf2,inf3,inf4,inf5,inf6} and also as a dark energy
candidate to describe the late-time acceleration phase
\cite{de00,de01,de0,de1,de2,de3,de4,de5,de6}. In \cite{bian1} it was found
that new Kasner-like solutions exist, while some cosmological solutions in
locally rotational spacetimes were derived in \cite{bian2,bian3}. Some effects
on the Mixmaster universe can be found in \cite{bian4,bian5}. In general, the various implications of $f(R)$-gravity - and of other theories of gravitation as well - from a cosmological perspective can be seen in \cite{Odnew1,Odnew2}. Other studies on
gravitational collapse can be encountered in \cite{col1,col2}, while static
spherically solutions were derived in \cite{staa0,staa1,staa2} with or without
a constant Ricci scalar. Black hole solutions have been also investigated in
the literature, for instance see \cite{bh00,bh03,bh00a,bh01,bh02,bh04,bh05,bh06} and
references therein, as also physical phenomena like, binary black hole merge
\cite{bhaa1}, anti-evaporation \cite{bhaa2}, black hole thermodynamics
\cite{akbar,farr} and many other.

It is well known that black string solutions can be easily constructed in
Einstein's gravity by trivially embedding black hole solutions in higher
dimensions. The same is also true for certain classes of modified theories of
gravitation (for example it holds for a certain type of Lovelock Lagrangians
\cite{Kastor,Giribet}). In many cases, numerical solutions have been presented in the
literature \cite{Wiseman,Kobayashi,Kudoh}. However, exact solutions are always
of special interest and there exists an extended bibliography over the subject
covering a great number of gravitational configurations: from three
dimensional charged black strings \cite{Horne}, to cosmological constant
solutions in an arbitrary number of dimensions \cite{Adolfo}, even in the
presence of axionic scalar fields \cite{Adolfo2}. Some exact solutions which
describe rotating black strings in $f\left(  R\right)  $-gravity in the
presence of a electromagnetic field were derived in \cite{stri2}, while some
asymptotic black strings solutions can be found in \cite{stri1} and a cosmic
sting solution in four dimensions in the context of scalar tensor theory has
been given in \cite{stri3}. The stability of black string solutions is always
an issue, since in general these geometries are unstable, see for example
\cite{Gregory,Alex1,Alex2}. However, counterexamples of this general rule
exist and stable solutions may also arise \cite{stable1,stable2}. In what regards other interesting gravitational solutions, a toroidal black hole has also emerged in the Einstein nonlinear sigma model \cite{AlexFab}. What is more, in the literature one can also find brane solutions in higher dimensional $f(R)$-gravity \cite{Chakra1,Chakra2}

In this work we start by investigating analytical solutions of power law
$f\left(  R\right)  $-gravity in a four-dimensional static spacetime. In this
context we derive the general analytical solution and try to see under which
conditions interesting gravitational objects may be described by it. We find
that a black brane/string solution can be distinguished for certain values of
the involved parameters in the case where the metric is axisymmetric. The
outline of the paper is as follows: In Section \ref{themodel}, we briefly
discuss $f\left(  R\right)  $-gravity and derive the field equations for the
spacetime of our consideration. In Section \ref{gensolution}, we obtain the
general solution for the induced system of equations. Section \ref{solution}
includes the main results of our analysis which is the black brane/string
solution of the field equations for the power-law theory $f(R)=R^{k}$. In
Section \ref{thermo}, we calculate the surface gravity and thus the
temperature of the system as well as the entropy on the horizon. Finally, in
Section \ref{discus}, we discuss our results and draw our conclusions.

\section{Preliminaries}

\label{themodel}

The action integral of $f\left(  R\right)  $-gravity constitutes a
modification of the Einstein-Hilbert action and is given by the following
expression%
\begin{equation}
\mathcal{S}=\int dx^{4}\sqrt{-g}f\left(  R\right)  , \label{ac.01}%
\end{equation}
where $R$ is the Ricci scalar constructed from the spacetime metric $g_{\mu
\nu}$. It follows from (\ref{ac.01}) that the field equations of Einstein's GR
are recovered when $f\left(  R\right)  $ is a linear function of $R$.

Variation with respect to the metric tensor gives the field equations%
\begin{equation}
f_{,R}R_{\mu\nu}-\frac{1}{2}fg_{\mu\nu}-\left(  \nabla_{\mu}\nabla_{\nu
}-g_{\mu\nu}\nabla_{\sigma}\nabla^{\sigma}\right)  f_{,R}=0, \label{ac.02}%
\end{equation}
where we have assumed that we are in vacuum, i. e. there does not exist any
matter source. The Ricci scalar contains second order derivatives of the
coefficients of the metric $g_{\mu\nu}$, hence, relation (\ref{ac.02})
provides a fourth-order system of differential equations.

An alternative way to write the latter is by the use of Einstein's tensor
together with the definition of an energy-momentum tensor of geometric origin.
In particular, we can rewrite (\ref{ac.02}) as
\begin{equation}
R_{\mu\nu}-\frac{1}{2}Rg_{\mu\nu}=k_{eff}T_{\mu\nu}^{eff}, \label{ac.03}%
\end{equation}
where $T_{\mu\nu}^{eff}$ is the effective energy momentum tensor that includes
the terms which make the theory deviate from General Relativity,
\begin{equation}
T_{\mu\nu}^{eff}=\left(  \nabla_{\mu}\nabla_{\nu}-g_{\mu\nu}\nabla_{\sigma
}\nabla^{\sigma}\right)  f_{,R}+\frac{1}{2}\left(  f-Rf_{,R}\right)  g_{\mu
\nu} \label{ac.04}%
\end{equation}
while $k_{eff}=\left(  f_{,R}\left(  R\right)  \right)  ^{-1}$ is a varying
gravitational constant. From the latter it follows that the theory is defined
in the Jordan frame.

Apart from the limit in which $f_{,RR}\left(  R\right)  =0$, where General
Relativity is recovered, it can be observed that any constant Ricci curvature
($R=R_{0}$) solution of General Relativity also satisfies field equations
(\ref{ac.03}) if the following algebraic condition relating the free
parameters of the $f\left(  R\right)  $ function to $R_{0}$ \cite{barot}
holds
\begin{equation}
2 f\left(  R_{0}\right)  -R_{0}f_{,R}\left(  R_{0}\right)  =0. \label{ac.05}%
\end{equation}

In our work we assume the following static metric with line element%
\begin{equation}
ds^{2}=-a(r)^{2}dt^{2}+N(r)^{2}dr^{2}+b(r)^{2} d\phi^{2}+ c(r)^{2} d\zeta^{2}
\label{genline}%
\end{equation}
It is interesting to note that, when $b(r)=c(r)$ and the line element is
invariant under rotations in the $\zeta-\phi$ plane, the general solution of a
geometry characterized by a constant Ricci scalar $R_{0}\neq0$ is
\begin{equation}
\label{metRconst}ds^{2} = \mp\left(  r^{2} -\frac{1}{r}\right)  dt^{2} +
\frac{12}{R_{0}} \left(  \frac{1}{r}-r^{2}\right)  ^{-1} dr^{2} + r^{2}
\left(  d\phi^{2} + d\zeta^{2}\right)  ,
\end{equation}
which - as we noted earlier - is also a solution of GR in the presence of a
cosmological constant. Line element \eqref{metRconst} can be also seen to be a
special case of a more general solution presented in \cite{Lemos} including
also an electromagnetic field. Clearly, the minus branch of \eqref{metRconst}
characterizes a black brane or string (depending on the topology of the ($\phi,
\zeta$) surface) when $R_{0}<0$. For $f(R)$-gravity the constant scalar
curvature is related to the parameters of the relative model through algebraic
equation \eqref{ac.05}.

The existence of a solution like \eqref{metRconst} is our motive to start
investigating a more general setting given by line element \eqref{genline}.
Thus, we begin by considering $b(r)\neq c(r)$ and turn to the general case
where $R$ is not a constant. The Ricci scalar is calculated to be\footnote{The
prime \textquotedblleft$^{\prime}$\textquotedblright\ denotes a total
derivative with respect to the variable $r$, that is $a^{\prime}\left(
r\right)  =\frac{da\left(  r\right)  }{dr}$.}
\begin{equation}
R=-\frac{2}{N^{2}}\left(  \frac{a^{\prime\prime}}{a}+\frac{b^{\prime\prime}%
}{b}+ \frac{c^{\prime\prime}}{c}+ \frac{a^{\prime}b^{\prime}}{ab}%
+\frac{a^{\prime}c^{\prime}}{ac}+ \frac{b^{\prime}c^{\prime}}{bc}+\right)  +2
\frac{N^{\prime}}{N^{3}}\left(  \frac{a^{\prime}}{a}+\frac{b^{\prime}}{b} +
\frac{c^{\prime}}{c}\right)  . \label{ac.06a}%
\end{equation}
and the gravitational field equations (\ref{ac.02}) are
\begin{equation}
R^{\prime2} f_{,RRR} + \left[  \left(  \frac{b^{\prime}}{b}+ \frac{c^{\prime}%
}{c}-\frac{N^{\prime}}{N}\right)  R^{\prime}+ R^{\prime\prime}\right]  f_{,RR}
- \left(  \frac{a^{\prime\prime}}{a} + \frac{a^{\prime} b^{\prime}}{a b} +
\frac{a^{\prime} c^{\prime}}{a c} - \frac{a^{\prime} N^{\prime}}{a N}\right)
f_{,R} -\frac{1}{2} N^{2} f = 0 \label{fe.01}%
\end{equation}%
\begin{equation}
R^{\prime} \left(  \frac{a^{\prime}}{a} +\frac{b^{\prime} }{b} +
\frac{c^{\prime} }{c}\right)  f_{,RR} - \left(  \frac{a^{\prime\prime}}{a} +
\frac{b^{\prime\prime}}{b} + \frac{c^{\prime\prime}}{c} - \frac{a^{\prime}
N^{\prime}}{a N} - \frac{b^{\prime} N^{\prime}}{b N} - \frac{c^{\prime}
N^{\prime}}{c N} \right)  f_{,R} -\frac{1}{2} N^{2} f =0 \label{fe.02}%
\end{equation}%
\begin{equation}
R^{\prime2} f_{,RRR} + \left[  \left(  \frac{a^{\prime}}{a}+ \frac{c^{\prime}%
}{c}-\frac{N^{\prime}}{N}\right)  R^{\prime}+ R^{\prime\prime}\right]  f_{,RR}
- \left(  \frac{b^{\prime\prime}}{b} + \frac{a^{\prime} b^{\prime}}{a b} +
\frac{b^{\prime} c^{\prime}}{b c} - \frac{b^{\prime} N^{\prime}}{b N}\right)
f_{,R} -\frac{1}{2} N^{2} f = 0 \label{fe.03}%
\end{equation}
\begin{equation}
R^{\prime2} f_{,RRR} + \left[  \left(  \frac{a^{\prime}}{a}+ \frac{b^{\prime}%
}{b}-\frac{N^{\prime}}{N}\right)  R^{\prime}+ R^{\prime\prime}\right]  f_{,RR}
- \left(  \frac{c^{\prime\prime}}{c} + \frac{a^{\prime} c^{\prime}}{a c} +
\frac{b^{\prime} c^{\prime}}{b c} - \frac{c^{\prime} N^{\prime}}{c N}\right)
f_{,R} -\frac{1}{2} N^{2} f = 0 \label{fe.04}%
\end{equation}

An alternative way to derive the field equations (\ref{fe.01})-(\ref{fe.03}),
together with \eqref{ac.06a} for the scalar curvature, is with the use of a
point-like Lagrangian in the minisuperspace approach. In particular, the
action of the Euler-Lagrange vector over the Lagrangian
\begin{equation}
\label{Lanc}L = \frac{2}{N} \left[  f_{R} \left(  c a^{\prime}b^{\prime}+a
b^{\prime}c^{\prime}+ b a^{\prime}c^{\prime}\right)  +f_{RR}\left(  b c
a^{\prime}R^{\prime}+ a c b^{\prime}R^{\prime}+ a b c^{\prime}R^{\prime
}\right)  \right]  + N a b c \left(  f - R f_{R}\right)
\end{equation}
with respect to the variables~$\left\{  N,a,b,c,R\right\}  $, produces a
system of equations equivalent to (\ref{ac.06a})-(\ref{fe.04}). It is
important to mention that (\ref{Lanc}) is a singular Lagrangian, due to the
presence of the degree of freedom $N$ for which no corresponding velocity
appears. In this case $N$ plays a role similar to that of the lapse function
in cosmological Lagrangians, the difference being that the evolution of the
present system is in the $r$ variable.

\section{The analytic solution}

\label{gensolution}

The method of selecting some special form for the metric so as to determine
the function $f\left(  R\right)  $ afterwards by construction is generally
problematic and lacks physical information. So, in this work, we prefer to
work in the inverse direction by first proving the integrability of the field
equations; that is, the existence of an actual solution for the theory, and
afterwards to try and derive a closed-form expression for it. In order to
prove the integrability of the system we need to determine conservation laws
that will help us reduce the order of the dynamical equations (\ref{ac.06a}%
)-(\ref{fe.04}). Among the different ways for the determination of
conservation laws the symmetry method and the singularity analysis have been
extensively applied in $f\left(  R\right)  $-gravity in several cosmological
studies \cite{sym1,sym2,sym3,sym4}.

As we discussed above a well-known solution admitted by the system
(\ref{ac.06a})-(\ref{fe.04}) is that of General Relativity with a cosmological
constant. However, that is only a particular solution and it is in general
unstable, since the stability conditions depend on the form of $f\left(
R\right)  $. By the term particular solution we mean that there are present
fewer integration constants due to the fact that there is a difference on the
physical degrees of freedom between General Relativity and $f\left(  R\right)
$ (with $f_{RR}\neq0$) gravity.

We choose to work with the symmetry method and in order to simplify the
problem we redefine the lapse function
\begin{equation}
\label{lapNc}N = \frac{n}{a b c \left(  R f_{R}- f\right)  },
\end{equation}
which leads to the equivalent Lagrangian that is of the form
\begin{equation}
L\left(  n,q^{\alpha},q^{\prime\alpha}\right)  =\frac{1}{n}G_{\alpha\beta
}q^{\prime\alpha}q^{\prime\beta}-n, \label{Lag2}%
\end{equation}
where $q=(a,b,c,R)$ and $G_{\mu\nu}$ is
\begin{equation}
\label{minmetc}G_{\mu\nu} = \left(  2 a b c \left(  R f_{R}-f\right)  f_{R}
\right)
\begin{pmatrix}
0 & c & b & b c \frac{f_{RR}}{f_{R}}\\
c & 0 & a & a c \frac{f_{RR}}{f_{R}}\\
b & a & 0 & a b \frac{f_{RR}}{f_{R}}\\
b c \frac{f_{RR}}{f_{R}} & a c \frac{f_{RR}}{f_{R}} & a b \frac{f_{RR}}{f_{R}}
& 0
\end{pmatrix}
.
\end{equation}

It has been shown in \cite{tchris1} that linear in the momenta conserved
quantities of the original constrained Lagrangian can be constructed by
Killing vector fields of $G_{\alpha\beta}$. The latter is the singular system
realization of the Jacobi-Eisenhart metric used in Newtonian mechanics. The
idea behind its utilization is to map solutions of a given energy to geodesic
flows on Riemannian spaces \cite{Arnold} (for applications and examples in
pseudo-Riemannian spaces see \cite{ein1,ein2}).

The mini-superspace is four dimensional and its metric $G_{\mu\nu}$ is
conformally flat. The simple transformation $(a,b,c,R)\rightarrow(x,y,z,w)$
with
\begin{equation}
a=e^{\frac{(3-6k)x+\sqrt{2}(1-2k)y+(k+1)z}{12k-6}},\;b=e^{\frac{1}{12}\left(
\frac{9(k-1)w}{k-2}+\frac{2(4k-5)z}{2k-1}+\sqrt{2}y\right)  },\;c=e^{\frac
{(6k-3)x+\sqrt{2}(1-2k)y+(k+1)z}{12k-6}},\;R=e^{\frac{\sqrt{2}(k-2)y-(k+1)w}%
{4k^{2}-10k+4}},
\end{equation}
leads $G_{\mu\nu}$, as given by \eqref{minmetc}, to become
\begin{equation}
G_{\mu\nu}=(k-1)ke^{w+z}\mathrm{diag}\{-1,\frac{1-k}{2k-1},\frac{k^{2}%
-1}{(1-2k)^{2}},-\frac{3(k-1)^{2}(k+1)}{2(k-2)^{2}(2k-1)}\}. \label{minmetc2}%
\end{equation}
The minisuperspace admits six Killing fields, which in these coordinates are
\begin{equation}%
\begin{split}
&  \xi_{1}=\partial_{x},\quad\xi_{2}=\partial_{y},\quad\xi_{3}=\partial
_{z}-\partial_{w},\quad\xi_{4}=y\partial_{x}+\frac{(1-2k)x}{k-1}\partial_{y}\\
&  \xi_{5}=\frac{3\left(  2k^{2}-3k+1\right)  w}{2(k-2)^{2}}+z\partial
_{x}+\frac{(1-2k)^{2}x}{k^{2}-1}\partial_{z}-\frac{(1-2k)^{2}x}{k^{2}%
-1}\partial_{w},\\
&  \xi_{6}=-\frac{(2k-1)\left(  \left(  6k^{2}-9k+3\right)  w+2(k-2)^{2}%
z\right)  }{2(k-2)^{2}}\partial_{y}-\frac{(1-2k)^{2}y}{k+1}\partial_{z}%
+\frac{(1-2k)^{2}y}{k+1}\partial_{w}.
\end{split}
\end{equation}
As is well known, they can be used to construct the conserved quantities of
the form $Q_{i}=\xi^{\alpha}\frac{\partial L}{\partial q^{\prime\alpha}}$.
With the help of first order relations like $Q_{i}=$const. it is easy to
derive the general solution for a generic $k\neq1$. Of course, as we see from
\eqref{minmetc2}, the cases $k=1/2$, $k=5/4$ and $k=2$ have to be treated
separately. The general solutions of these particular cases are given later in
the appendix. By having excluded the aforementioned values, the final result
in the original coordinates (after absorbing some unnecessary constants of
integration reads):
\begin{subequations}
\label{valuec}%
\begin{align}
a(r)  &  =e^{(\alpha+\gamma)r}(\cosh r)^{\frac{k-1}{5-4k}}\label{ref1}\\
N(r)  &  =C\,e^{\frac{\beta+2\alpha}{2k-1}r}(\cosh r)^{\frac{(8k-7)(k-1)}%
{(2k-1)(5-4k)}}\label{ref2}\\
b(r)  &  =e^{\beta r}(\cosh r)^{\frac{k-1}{5-4k}}\label{ref3}\\
c(r)  &  =e^{(\alpha-\gamma)r}(\cosh r)^{\frac{k-1}{5-4k}} \label{ref4}%
\end{align}
with $C$, $\alpha$, $\beta$ and $\gamma$ being the remaining constants of
integration. Of the latter three only two are actually independent, say
$\alpha$ and $\gamma$, then it can be seen that (\ref{ref1})-(\ref{ref4}) is a
solution under the condition
\end{subequations}
\begin{equation}
\beta=-\frac{\left\vert k-2\right\vert \sqrt{\frac{\alpha^{2}(2k-1)(5-4k)^{2}%
+2(k-1)\left(  \gamma^{2}\left(  8k^{2}-14k+5\right)  -3(k-1)^{2}\right)
}{5-4k}}+\alpha(k-2)(2k-3)}{2(k-2)(k-1)} \label{betaconst}%
\end{equation}
A first important remark that we can make by studying (\ref{ref1}%
)-(\ref{ref4}) is that, a typical uniform black string solution (i.e. one with
$c(r)=$const.) is not possible in this setting. For the latter to happen we
should require $k=1$, which corresponds to GR and it is a value excluded from
this analysis. The scalar curvature that corresponds to this solution is
\begin{equation}
R=\frac{6(k-1)ke^{-\frac{2(\beta+2\alpha)}{2k-1}r}}{C^{2}(2k-1)(4k-5)}\left[
\cosh(r)\right]  ^{\frac{2(2-k)}{(2k-1)(4k-5)}}.
\end{equation}

At this point we can formulate the line element in such a way so that we can
identify $b(r)$ as a radial type of variable. If we, for example, choose the
parameter $\beta$ to be
\begin{equation}
\beta=\pm\frac{k-1}{5-4k}, \label{betavalue}%
\end{equation}
then $b(r)$ can become a radial variable $\tilde{r}$ in the following manner:
By performing the transformation
\begin{equation}
r=\pm\frac{1}{2}\ln\left(  \tilde{r}^{\frac{5-4k}{k-1}}-1\right)  ,
\label{transfrad}%
\end{equation}
we get $b(\tilde{r})=2^{\frac{k-1}{4k-5}}\tilde{r}$, with the constant value
$2^{\frac{k-1}{4k-5}}$ being irrelevant since it can be absorbed in the metric
with a scaling transformation. Under a transformation like \eqref{transfrad},
the quantities $a$, $b$, $c$ and $Ndr$ transform as scalars. After a few
reparametrizations and scalings over all of the variables, we can re-write the
ensuing line-element as (for simplicity instead of $\tilde{r}$ we write $r$
understanding that it is a different variable than the one used in expressions
(\ref{ref1})-(\ref{ref4})):
\begin{equation} \label{gensolr}
\begin{split}
ds^{2}=  &  -r^{2}\left(  \frac{r^{\frac{5-4k}{k-1}}}{m}-1\right)
^{\frac{k-1}{4k-5}\pm\left(  \alpha+\gamma\right)  }dt^{2}+r^{\frac{2\left(
2k^{2}-2k-1\right)  }{(1-k)(2k-1)}}\left(  \frac{r^{-\frac{4k-5}{k-1}}}%
{m}-1\right)  ^{-\frac{8k^{2}-12k+2}{(1-2k)(5-4k)}\mp\frac{2\alpha}{1-2k}%
}dr^{2}\\
&  +r^{2}d\phi^{2}+r^{2}\left(  \frac{r^{\frac{5-4k}{k-1}}}{m}-1\right)
^{\frac{k-1}{4k-5}\pm\left(  \alpha-\gamma\right)  }d\zeta^{2},
\end{split}
\end{equation}
where the constant $m$ appearing in \eqref{gensolr} is associated with the initial $C$ that appears in \eqref{ref2}. Wherever a double sign appears, the upper corresponds to a solution for $k>2$,
while the lower for $k<2$. Of course, we always have to remember that not both
of $\alpha$ and $\gamma$ are free parameters, but bound through the condition
$\beta=\pm\frac{k-1}{5-4k}$, with $\beta$ being given by \eqref{betaconst}.
The properties of the resulting spacetime are related to the type of theory
that we study and which is characterized by the number $k$. The scalar
curvature for the metric at hand is
\begin{equation}
R=\frac{6k(4k-5)}{(k-1)(2k-1)}r^{\frac{2(k-2)}{(2k-1)(k-1)}}\left(
\frac{r^{\frac{5-4k}{k-1}}}{m}-1\right)  ^{\frac{2k-3}{(2k-1)(4k-5)}\mp
\frac{2\alpha}{2k-1}} \label{Ricciabc}%
\end{equation}
which indicates that, depending on the value of $k$, we can have curvature
singularities at $r=0$, $r=m^{\frac{1-k}{4k-5}}$ and at $r\rightarrow+\infty$.

As we can see from \eqref{Ricciabc}, the only way to ensure that the term
$\frac{r^{\frac{5-4 k}{k-1}}}{m}-1$ does not appear in the scalar curvature -
so that it does not lead to its divergence either at $r=m^{\frac{1-k}{4 k-5}}$
or at infinity - is by setting
\begin{equation}
\alpha= \pm\frac{3-2 k}{2(5-4k)}.
\end{equation}
This value inevitably results, through \eqref{betaconst}, in $\gamma= \pm
\frac{1}{2}$ that in its turn implies $b(r)=c(r)$. Thus, we are led in this
way to start considering the case where the metric admits a toroidal type of symmetry.

\section{The black brane solution}

\label{solution}

We proceed by considering the aforementioned special case where $b(r)=c(r)$.
This is of special interest since, as we are about to see, it produces a black
brane/string solution. The latter depends on how you interpret the $\phi$,
$\zeta$ variables and the type of topology with which you endow the
two-surface that they construct. We can either consider a topology
$S^{1}\times\mathbb{R}$ (cylinder), $S^{1}\times S^{1}$ (torus) or $\mathbb{R} \times \mathbb{R}$ (plane) which imply
that one, both or none (respectively) of $\phi$ and $\zeta$ are bounded and periodic variables.

Two procedures can be followed to extract the solution: we can either start
from the beginning, following a similar procedure like that of a previous
section and consider the resulting mini-superspace using ansatz
\begin{equation}
ds^{2}=-a(r)^{2}dt^{2}+N(r)^{2}dr^{2}+b(r)^{2}\left(  d\phi^{2}+d\zeta
^{2}\right)  \label{torline}%
\end{equation}
for the line element, or simply arrange the parameters in the general solution
(\ref{ref1})-(\ref{ref4}) so that $b(r)=c(r)$. Obviously, the latter happens
when $\gamma=\alpha-\beta$, which means that the resulting solution depends on
the parameters $C$, $\alpha$ and $\beta$, with the latter two related through
\eqref{betaconst}, when $\gamma=\alpha-\beta$ has been substituted in it.
Under these conditions it is easy to see that the general solution for the
$b(r)=c(r)$ case is
\begin{subequations}
\label{valuec2}%
\begin{align}
a(r)  &  =e^{(2\alpha-\beta)r}(\cosh r)^{\frac{k-1}{5-4k}}\label{tor1}\\
N(r)  &  =C\,e^{\frac{2\alpha+\beta}{2k-1}r}(\cosh r)^{\frac{(8k-7)(k-1)}%
{(2k-1)(5-4k)}}\label{tor2}\\
b(r)  &  =e^{\beta r}(\cosh r)^{\frac{k-1}{5-4k}}\label{tor3}\\
c(r)  &  =e^{\beta r}(\cosh r)^{\frac{k-1}{5-4k}} \label{tor4}%
\end{align}
with $\alpha$ and $\beta$ bound to satisfy the algebraic relation
\end{subequations}
\begin{equation}
40\alpha^{2}+8\alpha^{2}k(4k-9)+4\alpha\beta(5-4k)+\beta^{2}%
(4k-5)(4k-3)-3(k-2)k-3=0. \label{torbeta}%
\end{equation}

Once more we can choose to study a particular solution belonging in this
configuration. As in the previous section, we choose the parameter $\beta$ to
be given by \eqref{betavalue}. A value that helps us, with simple
transformation like \eqref{transfrad}, to associate $b(r)$ to a radial
distance, only this time we additionally have $c(r)=b(r)$. With the choice
\eqref{betavalue}, the constraint \eqref{torbeta} implies that
\begin{equation}
\alpha=\pm\frac{1-k}{4k-5}\quad\text{or}\quad\alpha=\pm\frac{3-2k}{2(5-4k)}.
\end{equation}
The first set of values leads to a solution of the form\footnote{Again, we understand that the $r$ variable appearing in the following line elements is different from the one in \eqref{valuec2}.}
\begin{equation}
ds^{2}=-r^{2}dt^{2}+r^{\frac{2(-2k^{2}+2k+1)}{(k-1)(2k-1)}}\left(
1-\frac{r^{\frac{5-4k}{k-1}}}{m}\right)  ^{\frac{2k}{1-2k}}dr^{2}+r^{2}\left(
d\theta^{2}+d\phi^{2}\right)  ,
\end{equation}
which describes a spacetime that, depending on $k$, can have its singularity
points at the origin $r=0$, at infinity and at $r=m^{\frac{1-k}{4k-5}}$. It is
interesting to note that just for the value $k=\frac{3}{2}$ the corresponding
geometry has a curvature singularity only at $r=0$, while $r=m^{\frac
{1-k}{4k-5}}$ is just a coordinate singularity. In all the other cases where
$r=m^{\frac{1-k}{4k-5}}$ is just a coordinate singularity, a curvature
singularity at $r\rightarrow\infty$ necessarily occurs.

The most interesting situation appears when $\alpha= \pm\frac{3-2 k}{2(5-4
k)}$. With this set of values, we can write the two following line elements
depending on the value of $k$:

\begin{enumerate}
\item For $1<k< \frac{5}{4}$
\begin{equation}
\label{linesol1}ds^{2} = -r^{2} \left(  \frac{r^{\frac{5-4 k}{k-1}}}%
{m}-1\right)  dt^{2} + r^{\frac{-4 k^{2}+4 k+2}{(k-1) (2 k-1)}} \left(
\frac{r^{\frac{5-4 k}{k-1}}}{m}-1\right)  ^{-1} dr^{2} + r^{2} \left(
d\phi^{2}+ d\zeta^{2}\right)  .
\end{equation}

\item For $k<1$ ($k\neq1/2$) or $k>\frac{5}{4}$
\begin{equation}
\label{linesol2}ds^{2} = -r^{2} \left(  1- \frac{r^{\frac{5-4 k}{k-1}}}%
{m}\right)  dt^{2} + r^{\frac{-4 k^{2}+4 k+2}{(k-1) (2 k-1)}} \left(  1-
\frac{r^{\frac{5-4 k}{k-1}}}{m}\right)  ^{-1} dr^{2} + r^{2} \left(  d\phi
^{2}+ d\zeta^{2}\right)  .
\end{equation}

\end{enumerate}

We make this distinction so that the sign of $g_{tt}$ (the coefficient of
$dt^{2}$ in the line element) is negative in each case when $r>m^{\frac{1-k}{4
k-5}}$. The two line elements are associated with constant, complex scalings
among the coordinates and the essential constant $m$ that appears in them.
Both of them are of course solutions to the field equations
\eqref{fe.01}-\eqref{fe.04} for an $f(R)=R^{k}$ theory.

The scalar curvature reads
\begin{equation}
\label{Ricsol}R = \pm\frac{6 k (4 k-5) r^{\frac{2 (k-2)}{2 k^{2}-3 k+1}}%
}{(k-1) (2 k-1)}%
\end{equation}
with the plus corresponding to the first line element, while the minus to the
second. In both situations the Kretschmann scalar $K=R^{\kappa\lambda\mu\nu
}R_{\kappa\lambda\mu\nu}$ is
\begin{equation}%
\begin{split}
K =  &  \frac{4}{(1-2 k)^{2} (k-1)^{2} m ^{2}} \Big[\left(  56 k^{4}-264
k^{3}+468 k^{2}-370 k+111\right)  r^{-\frac{2 \left(  8 k^{2}-16 k+9\right)
}{(2 k-1) (k-1)}}\\
&  -2 m (k-2)^{2} (2 k-1) r^{\frac{-8 k^{2}+18 k-13}{(2 k-1) (k-1)}} +3 m^{2}
\left(  8 k^{4}-20 k^{3}+13 k^{2}-2 k+2\right)  r^{\frac{4 (k-2)}{(k-1) (2
k-1)}}\Big] .
\end{split}
\end{equation}
Thus, it can be seen that for the range of values $k<1/2$ and $1<k\leq2$ the
space-time has a curvature singularity at $r=0$, while at the same time
exhibits a coordinate singularity at $r=r_{h}= m^{\frac{1-k}{4 k-5}}$ (as long
as $m>0$). We have to note here that, although we had initially excluded $k=2$
from the general solution, its consideration separately leads to the same
resulting spacetimes \eqref{linesol1} and \eqref{linesol2}, so we can include
it at this point as an admissible value for $k$.

Line element \eqref{linesol1} is appropriate for describing a (compactified) black
brane or string in vacuum $f(R)=R^{k}$ gravity for those theories for which
$1<k<\frac{5}{4}$. On the other hand, the spacetime characterized by
\eqref{linesol2} can be used for the same purpose when $k<\frac{1}{2}$ and
$\frac{5}{4}<k\leq2$. In both of these cases the surface $r=r_{h}=
m^{\frac{1-k}{4 k-5}}$ acts as an horizon ``hiding" the singularity. Particularly for the case of the toroidal topology we have to mention the work of Vanzo \cite{Vanzo} who was the first to consider such topological black holes. The
resulting spacetime is asymptotically Ricci flat, however the curvature
$R_{\kappa\lambda\mu\nu}$ does not tend to zero as $r\rightarrow\infty$. Thus, this geometry can not be smoothly deformed to any maximally symmetric space-time.

For the rest of values, that is $\frac{1}{2}<k<1$ and $k>2$, the curvature
singularity goes form the origin to infinity. Hence, we can state that, in the
theories corresponding to those values, such an object does not occur. We may
now proceed with studying the basic thermodynamic quantities for the
string and the compactified brane that appears in this context.

\section{Thermodynamic properties}

\label{thermo}

We continue by calculating basic thermodynamic quantities like the temperature
and the entropy for line elements \eqref{linesol1} and \eqref{linesol2} for
those values of $k$ for which they describe a black object. That is
$1<k<5/4$ for the first case and $k\in(-\infty,\frac{1}{2})\cup(\frac{5}%
{4},2]$ for the second. In both situations, we can perform a transformation
$t\mapsto\tau$ of the form
\begin{equation}
t=\tau\mp\int\sqrt{\frac{g_{rr}}{-g_{tt}}}dr \label{transtonull}%
\end{equation}
where $g_{rr}$ corresponds to the radial variable of the metric and thus,
express the latter in an ingoing Eddington - Finkelstein type of coordinate
system. The minus in transformation \eqref{transtonull} corresponds to line
element \eqref{linesol1}, while the plus is for \eqref{linesol2}. As a result,
we obtain
\begin{equation}
ds^{2}=\pm r^{2}\left(  1-\frac{r^{\frac{5-4k}{k-1}}}{m}\right)
dt^{2}+r^{\frac{2-k}{(k-1)(2k-1)}}dt dr + r^{2}\left(  d\phi^{2}+d\zeta
^{2}\right)
\end{equation}
with the $+$ referring to \eqref{linesol1} and the $1<k<\frac{5}{4}$ case,
while the minus is for \eqref{linesol2} and the rest of the values of $k$ as
discussed above.

In this coordinate system we may calculate the surface gravity $\kappa$ for
the given spacetimes, either from relation $\kappa= \Gamma^{t}_{tt}|_{r=r_{h}%
}$, where $\Gamma$ stands for the Christoffel symbols, or more formally by
\begin{equation}
\nabla_{\mu}(X^{\nu}X_{\nu})|_{r=r_{h}} = - 2 \kappa X_{\mu}|_{r=r_{h}},
\end{equation}
where $X^{\mu}=\frac{\partial}{\partial t}$ is the time-like Killing vector
field for which a Killing horizon exists at $r=r_{h}$. The resulting surface
gravity is given in terms of the essential constant $m$ as
\begin{equation}
\kappa=
\begin{cases}
-\frac{(4 k-5)}{2 (k-1)} m^{\frac{-2 k^{2}+2 k+1}{(2 k-1) (4 k-5)}}, & 1<k<
\frac{5}{4}\\
\frac{(4 k-5)}{2 (k-1)} m^{\frac{-2 k^{2}+2 k+1}{(2 k-1) (4 k-5)}}, & k
\in(-\infty, \frac{1}{2}) \cup(\frac{5}{4}, 2]
\end{cases}
\end{equation}
and allows us to easily deduce the Hawking temperature through the relation $T
= \frac{\kappa}{2\pi}$.

We know that the horizon entropy in $f(R)$ gravity is given in terms of the
first derivative of $f$ with respect to $R$ as calculated on the horizon,
i.e.
\begin{equation}
S \sim\sigma f_{R}|_{r=r_{h}},
\end{equation}
where $\sigma$ is the area of the horizon. In the case that we are dealing
with a string, we can use instead of $\sigma$ the area per unit length
$\tilde{\sigma} = 2 \pi r_{h}^{2} = 2 \pi m^{-\frac{2 (k-1)}{4 k-5}}$
\cite{Cai} and talk about an entropy density $\tilde{S}$, which we may
calculate to obtain
\begin{equation}
\tilde{S} \sim\tilde{\sigma} f_{R}|_{r=r_{h}} \sim%
\begin{cases}
k \left(  \frac{k (4 k-5)}{(k-1) (2 k-1)}\right)  ^{k-1} m^{-\frac{6
(k-1)^{2}}{(2 k-1) (4 k-5)}}, & 1<k< \frac{5}{4}\\
k \left(  \frac{(5-4 k) k}{2 k^{2}-3 k+1}\right)  ^{k-1} m^{-\frac{6
(k-1)^{2}}{(2 k-1) (4 k-5)}}, & k \in(-\infty, \frac{1}{2}) \cup(\frac{5}{4},
2].
\end{cases}
\end{equation}
From the form of $\tilde{S}$ we can see that the latter is positive only if
$0<k<\frac{1}{2}$ or when $k$ is a negative even integer. Thus, we can
distinguish a class of $f(R) = R^{k}$ theories - corresponding to these range
of values for $k$ - in which only solution \eqref{linesol2} results in a well
defined and positive entropy.

In the case where we consider a toroidal topology, so that the solution
represents a compactified black brane, none of the previous basic arguments changes, only
now we need to use, instead of $\tilde{\sigma}$, the area $\sigma= 4 \pi^{2}
r_{h}^{2}$, which implies that the resulting entropy $S$ is different from
$\tilde{S}$ up to a positive multiplicative constant.

Of course, for a complete thermodynamic description one should also take into account the mass/energy of the system. However, this would be a highly non-trivial calculation, since the space-time under consideration is not asymptotically maximally symmetric.

\section{Discussion}

\label{discus}

The main scope of this work is to find closed-form solutions for a four
dimensional static spacetimes in $f\left(  R\right)  $-gravity. We derived the
general solution for power-law $f(R)$-gravity with $f(R)=R^{k}$. In contrast
to General Relativity where black string solutions are allowed only in the
presence of the cosmological constant or matter, we saw that in $f\left(
R\right)  $-gravity, specifically in the $f\left(  R\right)  =R^{k}$ theory,
black string solutions are supported by the field equations for some ranges of
the power $k$ even in vacuum.

It is important to mention that solutions (\ref{linesol1}) and
\eqref{linesol2} reduce to \eqref{metRconst} with constant nonzero Ricci
scalar in the specific case when $k=2$. In the generic case of course, when
$k\neq2$, we can see from \eqref{Ricsol} that \eqref{linesol1} in its full
generality is not a constant Ricci scalar solution. However, (\ref{linesol1})
and \eqref{linesol2} are still only a special cases resulting from the most
general expressions (\ref{tor1})-(\ref{tor2}) with an appropriate fixing of
the integration constants.

We now continue with the discussion of the effective gravitational potential
for the spacetimes (\ref{linesol1}) and (\ref{linesol2}). The Hamiltonian of
the geodesic equations for the corresponding line element is%
\begin{equation}
\mp r^{2}\left(  \frac{r^{\frac{5-4k}{k-1}}}{m}-1\right)  \left(  \frac
{dt}{ds}\right)  ^{2} \pm r^{\frac{-4k^{2}+4k+2}{(k-1)(2k-1)}}\left(
\frac{r^{\frac{5-4k}{k-1}}}{m}-1\right)  ^{-1}\left(  \frac{dr}{ds}\right)
^{2}+r^{2}\left(  \left(  \frac{d\phi}{ds}\right)  ^{2}+\left(  \frac{d\zeta
}{ds}\right)  ^{2}\right)  =-\mu^{2} \label{ham1}%
\end{equation}
where $\mu^{2}=1$ for a timelike particle and $\mu^{2}=0$, for photons. Also,
wherever we have a double sign, the upper one corresponds to (\ref{linesol1}),
while the lower to (\ref{linesol2}).

The static axisymmetric spacetime (\ref{genline}) in general admits a four
dimensional Killing algebra consists by the symmetry vector $\partial_{t}$ and
the $E^{2}$ Lie algebra in the plane $\phi-\zeta$. The corresponding
conservation laws are%
\begin{equation}
\pm r^{2}\left(  \frac{r^{\frac{5-4k}{k-1}}}{m}-1\right)  \left(  \frac
{dt}{ds}\right)  =\mathcal{E}_{0}\mathcal{~~},~~r^{2}\left(  \frac{d\phi}%
{ds}\right)  =I_{\phi}~,~~r^{2}\left(  \frac{d\zeta}{ds}\right)  =I_{\zeta},
\end{equation}
and%
\begin{equation}
r^{2}\left(  \zeta\left(  \frac{d\phi}{ds}\right)  -\phi\left(  \frac{d\zeta
}{ds}\right)  \right)  =I_{\phi\zeta}.
\end{equation}

For simplicity let us consider from now on $m=1$, so that the coordinate
singularity rests at the radius $r=1$. With the use of the conservation laws
the Hamiltonian (\ref{ham1}) becomes%
\begin{equation}
\left(  \frac{dr}{ds}\right)  ^{2}=r^{-\frac{2\left(  4k^{2}-3k+2\right)
}{\left(  k-1\right)  \left(  2k-1\right)  }}\left[  \left(  \mathcal{E}%
_{0}^{2}\pm I_{0}^{2}\pm\mu^{2}r^{2}\right)  r^{\frac{4k}{k-1}}\mp r^{\frac
{5}{k-1}}\left(  I_{0}^{2}+\mu^{2}r^{2}\right)  \right]  \label{soo1}%
\end{equation}
where $I_{0}^{2}=I_{\phi}^{2}+I_{\zeta}^{2}$. The latter expression
\ corresponds to the Hamiltonian of classical particle of energy $\mathcal{E}$
with effective potential%
\begin{equation}
V_{eff}\left(  r\right)  =-\left(  r^{-\frac{2\left(  4k^{2}-3k+2\right)
}{\left(  k-1\right)  \left(  2k-1\right)  }}\left[  \left(  \mathcal{E}%
_{0}^{2}\pm I_{0}^{2}\pm\mu^{2}r^{2}\right)  r^{\frac{4k}{k-1}}\mp\left(
I_{0}^{2}+\mu^{2}r^{2}\right)  r^{\frac{5}{k-1}}\right]  +\mathcal{E}\right)
. \label{ham2}%
\end{equation}
Once more the upper signs correspond to the $1<k<\frac{5}{4}$ case, while the lower are for $k\in (-\infty,\frac{1}{2})\cup (\frac{5}{4},2]$. An important observation is that at the coordinate singularity, $r=1$, the
effective potential becomes%
\[
V_{eff}\left(  r\rightarrow1\right)  =-\left(  \mathcal{E}_{0}%
^{2}+\mathcal{E}\right)  .
\]

\begin{figure}[ptb]
\includegraphics[height=10cm]{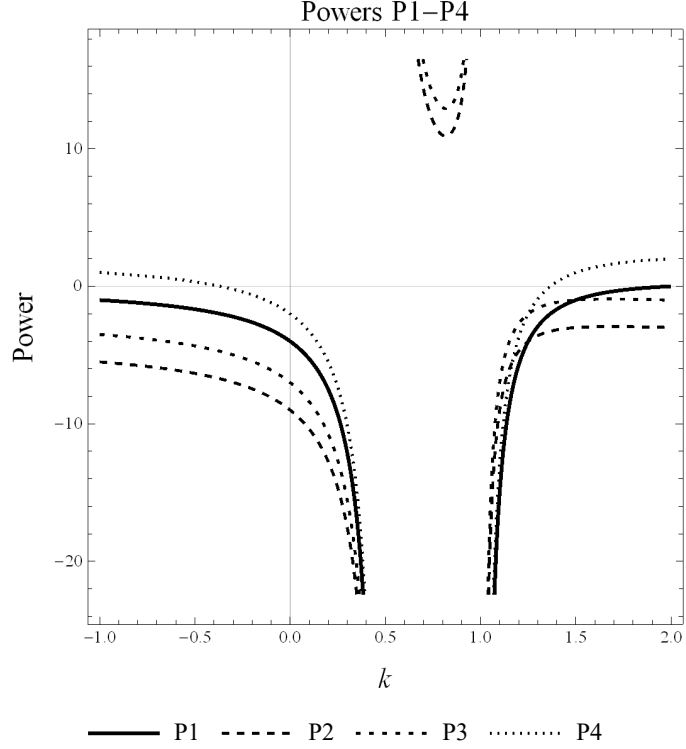}\centering\caption{Evolution of the
powers $P_{1-4}$ which lead the behaviour of the effective potential
(\ref{ham2}). When all the powers are negative the effective potential far
from the singularity becomes constant.}%
\label{veff}%
\end{figure}

\begin{figure}[ptb]
\includegraphics[height=7cm]{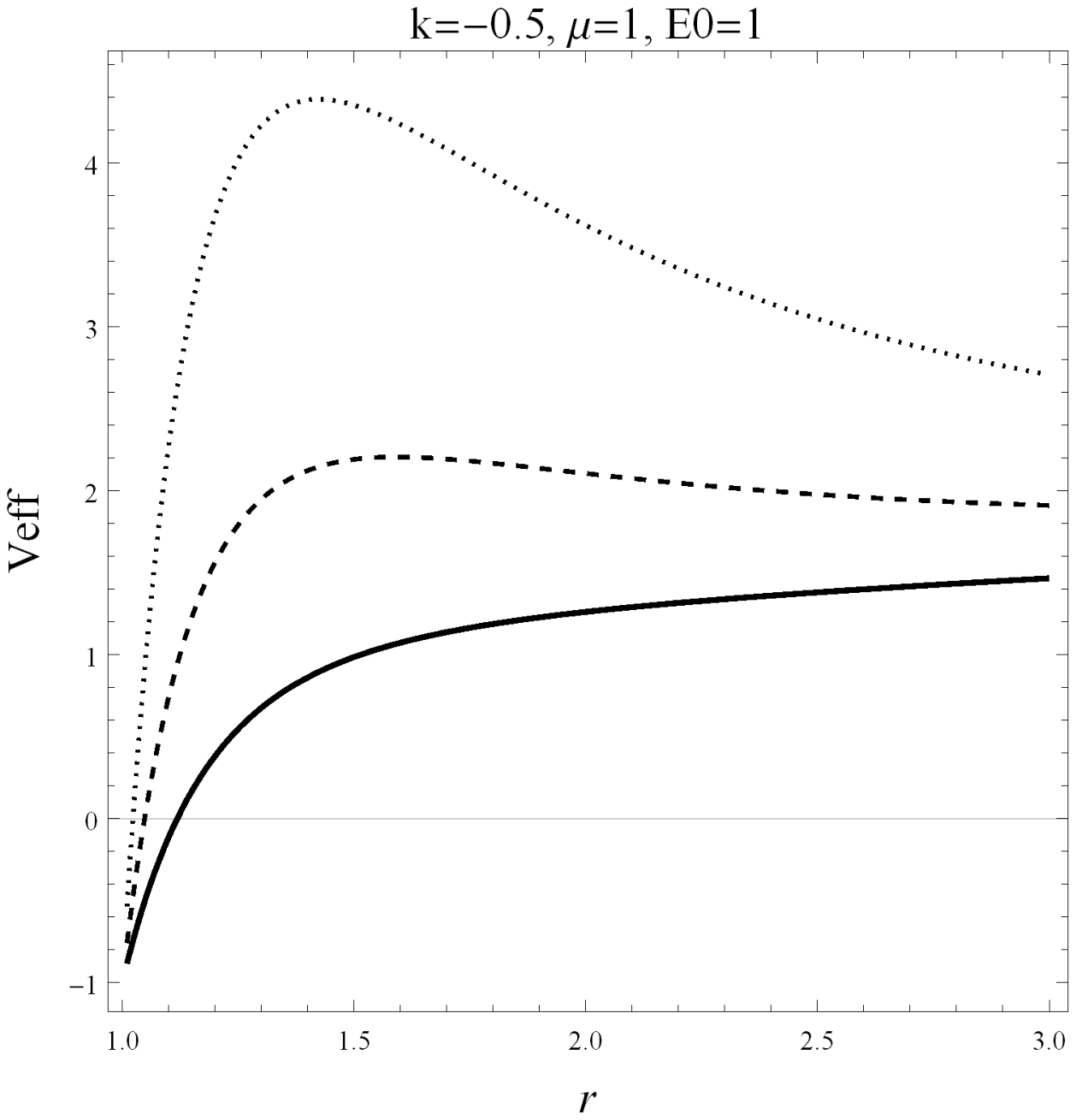}\centering\includegraphics[height=7cm]{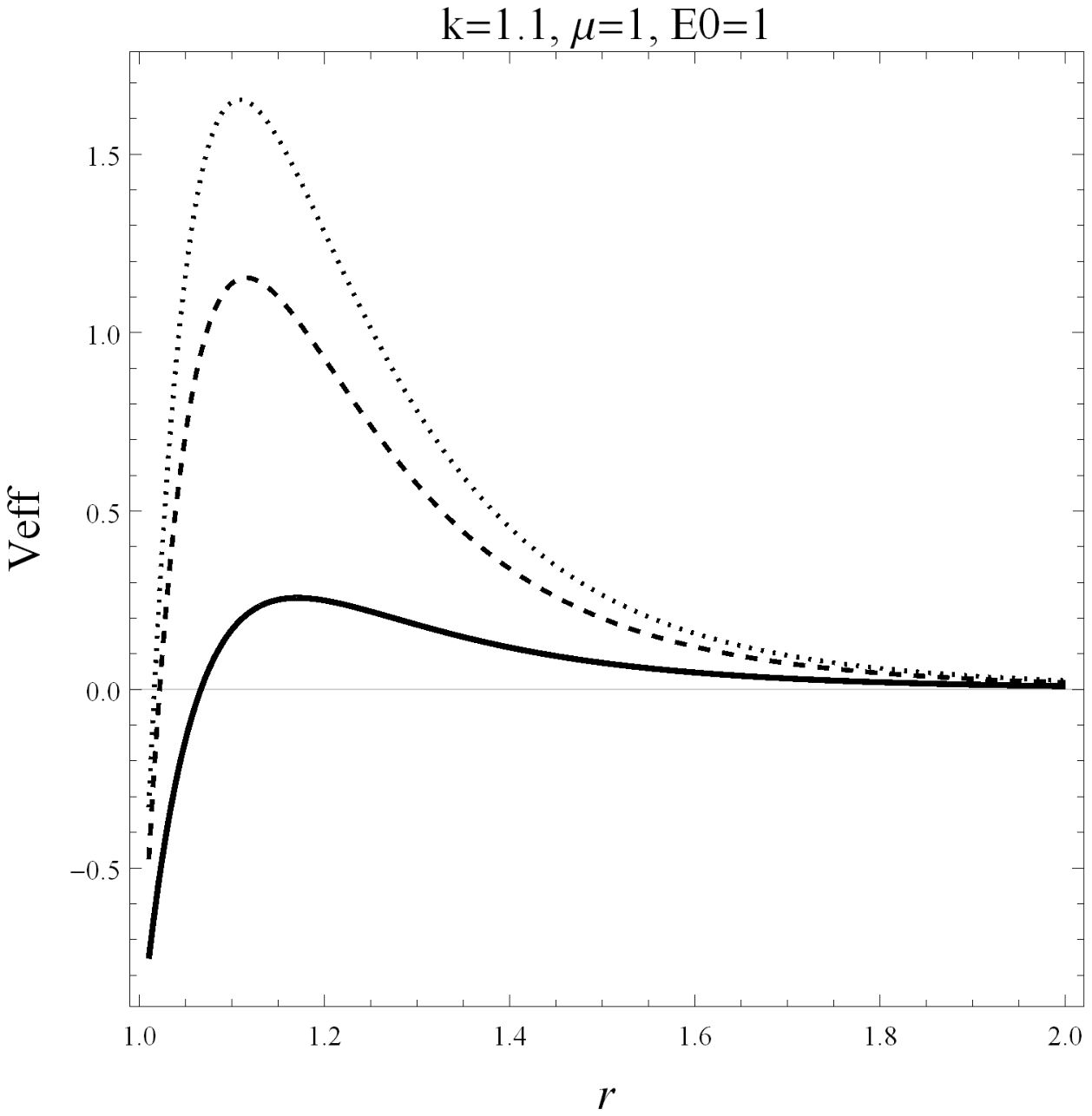}\centering\caption{Qualitative
evolution of the effective potential (\ref{ham2}) for timelike test particle,
$\mu^{2}=1$, $k=-0.5~$(Left Fig.) and $k=1.1~$(Right Fig.).~Solid line is for
$I_{0}^{2}=1,$ dashed line is for $I_{0}^{2}=2.5~$and dot-dot line is for
$I_{0}^{2}=3$. For simplicity we considered~$\mathcal{E=}0$.}%
\label{veff2}%
\end{figure}

\begin{figure}[ptb]
\includegraphics[height=7cm]{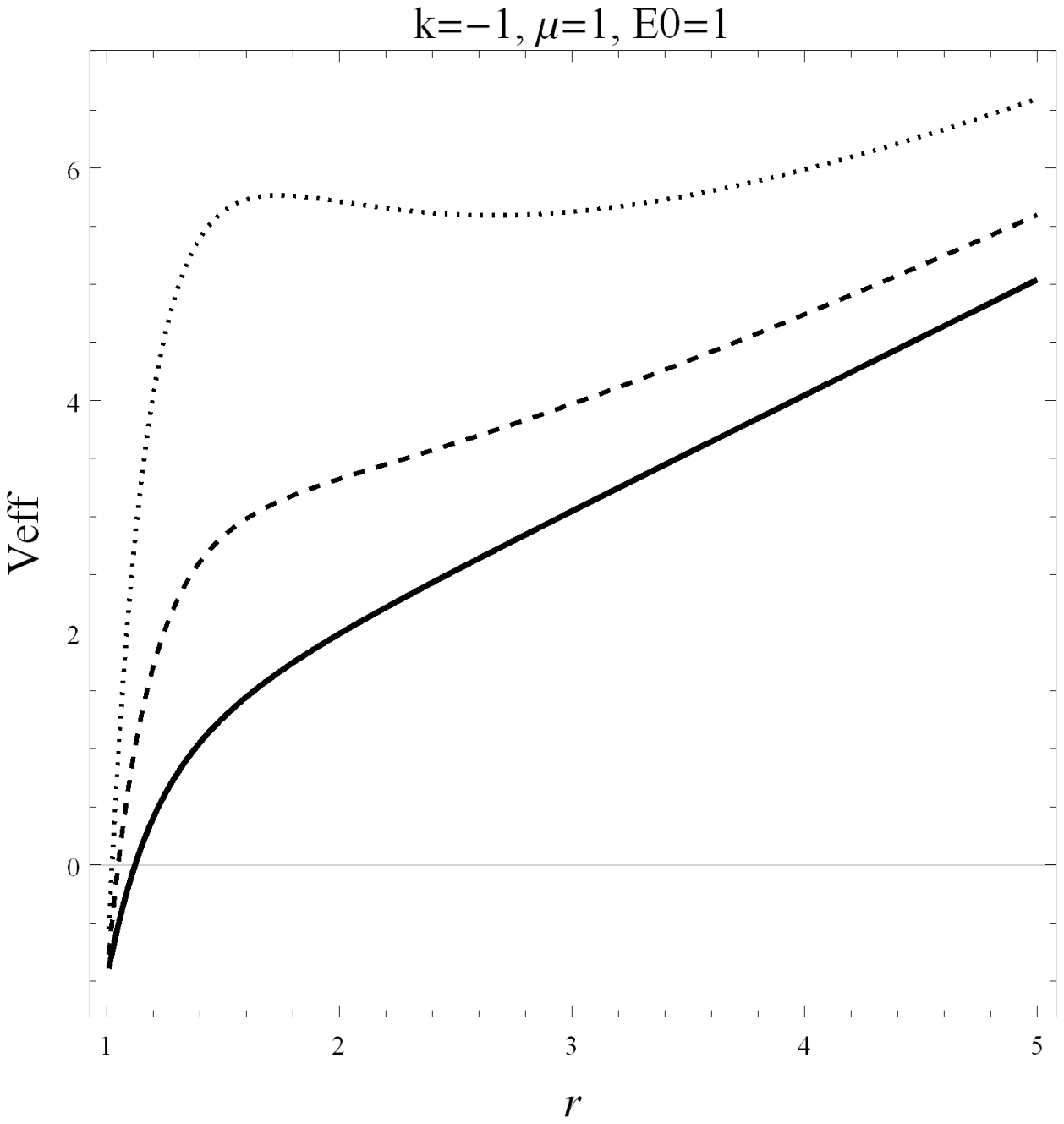}\centering\caption{Qualitative
evolution of the effective potential (\ref{ham2}) for timelike test particle,
$\mu^{2}=1$, $k=-0.5~$(Left Fig.) and $k=1.1~$(Right Fig.).~Solid line is for
$I_{0}^{2}=1,$ dashed line is for $I_{0}^{2}=2.5~$and dot-dot line is for
$I_{0}^{2}=3$. For simplicity we considered~$\mathcal{E=}0$.}%
\label{veff33}%
\end{figure}

From (\ref{ham2}) we can define four different powers which lead the behaviour
of $V_{eff}\left(  r\right)  $. Those are given by
\[
P_{1}=\frac{2\left(  k-2\right)  }{\left(  k-1\right)  \left(  2k-1\right)
}~,~P_{2}=-\frac{8k^{2}-16k+9}{\left(  k-1\right)  \left(  2k-1\right)  },
\]%
\[
P_{3}=-\frac{4k^{2}-10k+7}{\left(  k-1\right)  \left(  2k-1\right)
}~\text{and }P_{4}=\frac{2\left(  2k^{2}-2k-1\right)  }{\left(  k-1\right)
\left(  2k-1\right)  }.
\]
In Fig. \ref{veff} the evolution of the powers $P_{1-4}$ is presented where it
can be seen for which values of $k$ the effective potential $V_{eff}$ becomes
a constant far from the singularity. Furthermore, the qualitative evolution of
the effective potential $V_{eff}$ outside the horizon can be seen, for three
values of $k$, in Figs. \ref{veff2}, and \ref{veff33}.

Before we conclude, it is important to mention that another special solution
can be recovered from the general expressions (\ref{tor1})-(\ref{tor2}). It is
a power-law solution which is similar to that of the cosmological Bianchi type
I axisymmetric spacetime in $f(R)=R^{k}$, that leads to a Kasner-like universe
\cite{bian1}. Therefore, in a similar way more general solutions may be
constructed also in a cosmological context.

\begin{acknowledgments}
This work is financial supported by FONDECYT grant 1150246, CONICYT  DPI 20140053 project (AG) and FONDECYT grant 3160121
(AP). AP thanks the University of Athens for the hospitality provided while
part of this work carried out.
\end{acknowledgments}

\appendix

\section{Solutions when $k=2$, $k=5/4$ and $k=1/2$}

\label{appen}

For the sake of completeness we just state here the solutions that emerge for
the cases which we needed to exclude throughout the initial calculations, that
is $k=2$, $k=5/4$ and $k=1/2$. We state those solutions for the general
anisotropic line element \eqref{genline}. The case $c(r)=b(r)$ can be
recovered with a simple fixing of parameters. By working in a similar manner
as in the generic $k$ case, we obtain the following results:

\begin{itemize}
\item When $k=2$ the spacetime is characterized again by the functions
(\ref{ref1})-(\ref{ref4}), only now the algebraic condition among the constants
is not \eqref{betaconst} but rather
\begin{equation}
\beta=\frac{1}{2}\left(  \pm\sqrt{-9\alpha^{2}-6\gamma^{2}+2}-\alpha\right)  .
\end{equation}
The $b(r)=c(r)$ case can be recovered in the same manner as for the generic
$k$ solution, by setting $\gamma=\alpha-\beta$ in the corresponding expressions.

\item For $k=5/4$ the line element that satisfies the field equations reads
\begin{equation}
ds^{2} = - e^{\alpha r+\frac{e^{r}}{2}} dt^{2} + e^{\nu r+e^{r}} dr^{2} +
e^{\frac{1}{2} \left(  \beta r+e^{r}\right)  } d\phi^{2} + e^{\frac{1}{2}
\left(  e^{r}-r (2 \alpha+\beta-3 \nu+1)\right)  } d\zeta^{2}%
\end{equation}
together with the following algebraic condition among the constants appearing
in it
\begin{equation}
4 \alpha^{2}+2 \alpha(\beta-3 \nu+1)+\beta^{2}-3 \beta\nu+\beta+3 \nu^{2}-4
\nu+1 = 0 .
\end{equation}
Thus, the geometry has two essential constants. The case $b(r)=c(r)$ is
recovered when $\alpha= \frac{1}{2} (-2 \beta+3 \nu-1)$. One can easily
express this solution in a gauge where the coefficient $b(r)$ of $d\phi^{2}$
can be considered as a radial distance by performing a transformation
$r\rightarrow\tilde{r}$ that involves the Lambert $W$ function, $r = \frac{4
\ln(\tilde{r} )}{\beta}-W(\frac{\tilde{r}^{4/\beta}}{\beta})$.

\item On the other hand, for $k=1/2$ we get the solution
\begin{equation}
ds^{2}=-e^{\alpha r}dt^{2}+e^{e^{r (\alpha+\beta+\gamma)}+\nu r}%
dr^{2}+e^{\beta r} d\phi^{2}+ e^{\gamma r} d\zeta^{2},
\end{equation}
where the constants appearing in the solution need to satisfy
\begin{equation}
\alpha^{2}+(\gamma-\nu) (\alpha+\beta+\gamma)+\alpha\beta+\beta^{2} = 0 .
\end{equation}
Clearly, the $b(r)=c(r)$ solution corresponds to taking $\gamma= \beta$. By
performing the transformation $r = \frac{\ln(\tilde{r})}{\beta}$, which
corresponds to $b(\tilde{r})=\tilde{r}$, it can be seen that this spacetime
may exhibit a singularity at the origin $\tilde{r}=0$ or at infinity,
depending on the values of the free parameters.
\end{itemize}

\end{document}